# Dynamic analysis of refractive index evolution and diffraction properties during single-photon polymerization of photopolymers for micro-optical applications

**Andreas Heinrich, Manuel Rank**

Aalen University, Aalen School of Applied Photonics, Anton Huber Strasse 21, 73430 Aalen, Germany

**Abstract**. Photopolymerization enables the production of micro-optical elements, such as diffractive optical elements or GRIN optics. This process utilizes targeted spatial modulation of the refractive index, which is achieved through additive manufacturing. In this context, a thorough understanding of the dynamic processes during curing is essential in order to be able to accurately predict the optical function of the element. For this reason, this work investigates the kinetics and resulting optical properties of an acrylate-based photopolymer under UV irradiation using a DLP projection system. The experimental approach combines two measurement methods: On the one hand, the absolute change in the refractive index is determined in a time-resolved manner at the interface of a prism using a method based on total reflection. On the other hand, the formation of a phase grating in the volume of the polymer is monitored in real time by analyzing the diffraction orders of a coherent sample laser. The results show a characteristic S-shaped curve of the refractive index change, which reflects the phases of polymerization: oxygen inhibition, autoacceleration, and vitrification. Analysis of the diffraction patterns reveals complex intensity curves and substructures in the diffraction orders. These could be traced back experimentally and through simulations to the discrete pixel structure of the DLP projector and the existing dead zones. Furthermore, the simulation model developed on the basis of Fourier optics reproduces the experimental diffraction patterns and confirms the hypothesis that scattered light and radical diffusion lead to time-delayed polymerization in the theoretically unexposed areas. This results in a reduction of the refractive index contrast over time. This work thus provides parameters for the simulation and optimization of exposure strategies in 3D printing of micro-optics.

# 1    Introduction - photopolymerization

Photopolymerization is a fundamental process that describes the conversion of liquid monomer and oligomer mixtures into solid, cross-linked polymer networks through the targeted application of light energy. This can be used to create optically transparent microelements. In principle, a distinction can be made between 2-photon and 1-photon processes. In a 1-photon process, light from a UV-curing laser or UV projector is directed onto the polymer to polymerize it locally. If a projection system is used, a completely two-dimensional area can be cured in a single shot. Different irradiation intensities can also be selected for the individual pixels, so that locally different polymerization structures are created. This can be used, for example, to create diffractive optical elements in pixel resolution. A crucial point here is that the irradiance or dose has a direct influence on the locally generated refractive index. This enables the targeted realization of GRIN optics. However, a basic prerequisite for this is that the relationship between process parameters and refractive index development or optical effect can be described in combination with an understanding of the material.

## *1.1 Basic principles of light-induced polymerization*

Photopolymerization is a chemical reaction triggered by photons. The process begins with a liquid or viscous, photosensitive resin, which typically consists of three main components: monomers, oligomers, and a photoinitiator. Monomers are low-molecular-weight reactive molecules that serve as building blocks for the polymer chains. Oligomers are short, pre-polymerized chains that largely determine the basic properties of the final material, such as flexibility, hardness, chemical resistance, and optical properties. In 1-photon absorption, the photoinitiator usually absorbs light in the ultraviolet (UV) or visible range to start polymerization[1,2].

In principle, two polymerization mechanisms can be distinguished: free radical polymerization and cationic reaction. Free radical polymerization is initiated by free radicals and is typically used for the polymerization of monomers with acrylate or methacrylate groups. A major disadvantage of this mechanism is its sensitivity to atmospheric oxygen, which can scavenge the radicals and inhibit the reaction. This can lead to incomplete curing, especially on the surface. Cationic polymerization is initiated by cations (usually strong acids). It is mainly used for the ring-opening polymerization of epoxides and vinyl ethers. Cationic polymerization is generally slower than radical polymerization, but has the decisive advantage of being insensitive to oxygen. It also exhibits less volume shrinkage during curing, which leads to lower internal stresses.

In principle, photopolymerization proceeds in the following steps:
1. The photoinitiator molecules absorb the photon energy from a light source.
2. The absorbed energy turns the photoinitiator into an excited state and it decays or reacts to produce a highly reactive chemical species. These can be either free radicals (molecules with an unpaired electron: R·) or cations (positively charged ions).
3. The reactive species R· attacks the reactive functional groups (e.g., double bonds in acrylates) of the monomers M and oligomers, thus initiating a chain reaction (R-M·).
4. The reaction propagates as more and more monomer and oligomer units are added to the growing polymer chain. Since the monomers and oligomers used often have more than one reactive group per molecule, not only linear chains are formed, but also a three-dimensional, covalently bonded network via cross-links between the polymer chains. This cross-linking process is responsible for the conversion of the liquid resin into a solid plastic.
5. The reaction stops when two growing radical chains react with each other—either by combining (forming a single chain) or by disproportionation (hydrogen transfer).

In contrast to thermal polymerization, where polymerization is enabled by slow, diffuse heat propagation, photopolymerization allows for spatial and temporal control at high reaction speeds. This is particularly noteworthy in radical polymerization, which makes it attractive for 3D printing of micro-optics or diffractive optical elements. For this reason, it will be examined in more detail below.

*1.2 Chemical formulation of UV-curable resins*

In principle, resins for photopolymerization consist of oligomers and monomers[1,2]. Oligomers are molecules with a relatively high molecular weight that already consist of several monomer units and are often highly viscous. They form the basic structure of the polymer network and are the main carriers of the final material properties. Monomers are low-molecular compounds. On the one hand, they serve as reaction diluent. On the other hand, the monomers are completely incorporated into the network during polymerization and thus also contribute to the final properties. Unlike non-reactive solvents, they do not evaporate, which enables solvent-free formulations.

Due to their very high reactivity, the most important chemical classes for free-radical systems are acrylates and methacrylates. A decisive parameter here is functionality, i.e., the number of reactive groups per monomer or oligomer molecule. Monofunctional monomers (one reactive group) reduce viscosity most effectively, but result in lower crosslink density. Multifunctional monomers (two or more reactive groups) increase crosslink density. However, this is often accompanied by higher shrinkage.

The third component of UV-curable resins is the photoinitiator[3], which enables the reaction to be catalyzed by generating the reactive species that initiate polymerization. There are two types of photoinitiators. In type I photoinitiators (unimolecular), the molecules decay directly into two or more free radicals after light absorption through a unimolecular cleavage reaction (homolysis). This process is very fast and efficient. This class includes, for example, acylphosphine oxides such as TPO and BAPO, which absorb particularly in the near UV and violet range. Type II photoinitiators (bimolecular) do not directly generate radicals. After absorbing light, they transit into a long-lived excited triplet state. In this state, they react with a second molecule, a so-called co-initiator (e.g., tertiary amine or thiol). Hydrogen or electron transfer generates a radical on the co-initiator molecule, which then starts the polymerization. This two-step process is generally slower than type I initiation, so type II is not usually used for 1-photon polymerization.

In addition to these basic components, additives are often added to modify the resin properties[2,4-7], such as fillers to change the mechanical properties, photosensitizers to extend the spectral range of the resin, stabilizers and inhibitors to prevent unwanted polymerization, oxygen scavengers to consume the oxygen dissolved in the resin, or dyes to vary the optical properties.

*1.3 Challenges of radical polymerization*

One major challenge is oxygen inhibition. Molecular oxygen ($O_2$), which is dissolved in the resin and diffuses in from the atmosphere, is itself a diradical. It reacts extremely quickly with the initiator radicals or the growing polymer radicals and forms peroxy radicals (R-OO·). These peroxy radicals are very stable and not reactive enough to continue the polymerization chain efficiently, which effectively stops the reaction. This effect is strongest at the interface with air and leads to incomplete curing on the surface. To prevent this, one can work under an inert atmosphere (e.g., nitrogen), increase the light intensity or initiator concentration (to accelerate radical generation and "consume" the oxygen more quickly), or add oxygen scavengers to the resin[7,13].

*1.4 Photophysical and photochemical reactions*

The absorption of UV light for photochemical reactions is described in the simplest approach by Beer-Lambert's law. It applies to the dimensionless extinction A:

$$A = \log_{10}\left(\frac{I_0}{I}\right) = \epsilon c l, \qquad (1)$$

where $I_0$ is the intensity of the incident light beam, $I$ is the intensity of the transmitted light beam, $\epsilon$ is the extinction coefficient, $c$ is the concentration of the photoinitiator (PI), and $l$ is the layer thickness.

Since the light intensity decreases exponentially with the penetration depth into the resin, polymerization only takes place up to a depth where the remaining light energy is sufficient to generate a critical concentration of reactive species. This leads to a direct relationship between the cure depth $C_d$, the PI concentration, the molar absorption coefficient, and the light dose (intensity x time) applied. However, the cure depth is not a static quantity, but develops dynamically during the exposure process. As the PI absorbs light, it is consumed and converted into other molecules that often have lower absorption at the excitation wavelength (photobleaching). This leads to a decrease in the concentration $c$ of the absorbing species over time. This makes the resin increasingly transparent, allowing photons to penetrate deeper into the layer, i.e., the cure depth increases with increasing exposure time.

However, Beer-Lambert's law also has its limitations. Strictly speaking, it only applies to non-scattering, homogeneous media. In real photopolymers, however, light scattering occurs. The formation of the polymer network itself can also lead to inhomogeneities (e.g., due to local refractive index changes) and thus to scattering[8-10].

After a photon is absorbed by the photoinitiator molecule, a series of ultrafast photophysical processes begin, which can be described by a Jablonski diagram[11]:

1. The absorption of a photon lifts an electron from the ground electronic state (a singlet state, $S_0$) to a higher excited singlet state (e.g., $S_1$, $S_2$, etc.). This process takes place in the femtosecond range.
2. In higher excited states, the molecule relaxes very quickly (picoseconds) non-radiatively to the lowest excited singlet state ($S_1$), releasing excess energy to the environment as heat.

3. This $S_1$ state is relatively short-lived (nanoseconds) and either fluorescence occurs, i.e., the molecule can return to the ground state $S_0$ by emitting a photon (which is undesirable for photopolymerization). Or intersystem crossing takes place. In this case, the molecule changes its spin state and transitions to a slightly lower energy but significantly longer-lived excited triplet state ($T_1$). In this state, the spins of two electrons are parallel.
4. Due to its long lifetime (>> ps), the $T_1$ state has sufficient time to undergo the chemical reactions necessary for polymerization, i.e., in type I photoinitiators, the energy of the $T_1$ state is sufficient to cause homolytic cleavage of a weak covalent bond in the molecule, leading to the direct formation of two free radicals.

Finally, it should be noted that not every photon absorbed by the photoinitiator leads to the formation of a reactive species. The efficiency of this conversion process is described by the quantum yield ($\Phi$). It is defined as the ratio of the number of events that have taken place (e.g., decayed PI molecules) to the number of photons absorbed. In reality, $\Phi$ is always less than 1, as competing processes such as fluorescence or internal conversion, etc., reduce the efficiency of radical formation[12].

The overall initiation rate, i.e., the rate at which reactive species are generated, is thus a product of the rate of light absorption (determined by light intensity, PI concentration, and $\epsilon$) and the efficiency of this absorption (determined by $\Phi$).

The curing process can be described phenomenologically using so-called curing curves based on Jacobs' model[20]. It introduces a binary threshold concept for curing, i.e., the resin remains liquid until it has absorbed a specific amount of energy sufficient to generate enough free radicals to overcome inhibitors (primarily dissolved oxygen) and form a cross-linked gel network. This threshold is referred to as critical energy $E_c$. Curing only occurs where $E(z) > E_c$. The cure depth $C_d$ is defined as the specific depth $z$ at which the energy $E(z)$ corresponds exactly to the critical energy $E_c$. This results in:

$$E_c = E_0 exp\left(-\frac{C_d}{D_p}\right) \quad (2)$$

with $D_p$ as the penetration depth, which is given by

$$I(z) = I_0 exp\left(-\frac{z}{D_p}\right) \quad (3)$$

as the reciprocal of the absorption coefficient.

Ultimately, this results in the fundamental working curve equation for 1-photon polymerization:

$$C_d = D_p ln\left(\frac{E_0}{E_c}\right) \tag{4}$$

*1.5 Kinetics of photopolymerization*

The kinetics of photopolymerization describes how quickly monomers are converted into a polymer network and which factors influence this speed. This is particularly complex because the system changes from a low-viscosity liquid to a high-viscosity gel and finally to a glass-like solid. These physical changes have a direct effect on the chemical reaction rates.

A kinetic model of radical photopolymerization breaks down the overall process into the following steps or rates[8,12]:

1. The initiation rate $R_i$, at which the primary reactive radicals are generated, is directly proportional to the absorbed light intensity. This in turn depends on the incident light intensity ($I_0$), the concentration of the photoinitiator ($c_{PI}$), the molar absorption coefficient or extinction coefficient ($\epsilon$), and the quantum yield ($\Phi$) of the initiator:

   $$R_i = 2\Phi I_0 (1 - 10^{-\epsilon c_{PI} l}) \tag{5}$$

   The factor 2 takes into account that type I initiators often produce two radicals per molecule.

2. The propagation rate $R_p$ describes the rate of chain growth at which monomers are consumed. It is proportional to the concentration of monomers ($c_M$) and the total concentration of active polymer radicals ($c_{M\cdot}$) and is determined by the propagation rate constant ($k_p$):

   $$R_p = k_p c_M c_{M\cdot} \tag{6}$$

3. The termination rate $R_t$ describes the rate at which active radical chains are deactivated. In radical polymerization, this is typically a bimolecular reaction in which two radicals collide. Its rate is proportional to the square of the radical concentration and is described by the termination rate constant ($k_t$):

   $$R_t = k_t (c_{M\cdot})^2 \tag{7}$$

To solve these equations, a quasi-steady-state assumption can be made, i.e., after a very short start-up phase, an equilibrium is established in which the rate of radical generation is equal to the rate of radical termination ($R_i = R_t$). This makes it possible to eliminate the difficult-to-measure radical concentration $c_M$. and derive an equation for the total polymerization rate ($R_p$) that depends on measurable parameters.

One of the factors[8,10,13,14] influencing the polymerization rate or kinetics is light intensity ($I_0$): higher light intensity leads to a higher initiation rate and thus to faster overall polymerization. This can also lead to higher final conversion, as the reaction proceeds faster than the diffusion of inhibitors such as oxygen. However, excessive intensity can consume the photoinitiators on the surface too quickly (severe photobleaching), which can lead to uneven curing.

Another factor is the concentration of the photoinitiator. If this is increased, light absorption increases and thus the initiation rate. However, if the concentration is too high, the initiator itself acts as a light filter, absorbing the light in the uppermost layers and preventing sufficient curing.

The monomer concentration directly influences the propagation rate. Since polymerization is an exothermic reaction that releases heat, this can lead to a significant increase in temperature in the sample. A higher temperature as an additional factor generally increases molecular mobility and rate constants, which accelerates the reaction.

Viscosity and diffusion effects also represent a complex aspect. During polymerization, the viscosity increases by many orders of magnitude. This transition from liquid to solid (gelation and vitrification) severely restricts the mobility of all molecules and leads to diffusion-controlled effects. One example of this is autoacceleration (Tromdsdorff-Norrish effect). In a certain conversion range, the diffusion of large, heavy polymer radicals is more severely impeded than that of small, mobile monomer molecules. This means that the termination reaction (the collision of two large chains) is clearly slowed down, while the propagation reaction (a small monomer diffuses to the end of a chain) still proceeds relatively quickly. The result is an increase in the concentration of free radicals and a dramatic acceleration of the polymerization rate. However, at very high conversion, the system reaches the glass transition point and autodeceleration occurs. The material becomes so stiff that even the diffusion of the small monomer molecules is severely restricted. The propagation reaction slows down dramatically. As a result, the reaction often does not proceed completely and a certain proportion of unreacted functional groups remain trapped in the network.

Overall, this results in a characteristic sigmoid-shaped or S-shaped curve when the conversion is plotted against time. It is a macroscopic signature of these complex processes. The initial flat region (induction period) reflects the inhibition by oxygen. The steep rise of the curve shows the rapid propagation phase, often amplified by autoacceleration. The final plateau, where the reaction flattens out and a final, often incomplete conversion is achieved, is the result of vitrification, in which the glassy state freezes molecular motion.

*1.6 Polymerization and refractive index*

The polymerization process, typically a radical chain reaction, breaks the $\pi$ bonds (e.g., C=C double bonds in acrylates) and forms new, extremely strong and short covalent $\sigma$ bonds between the monomer units. The crucial point is the exchange of distances: the relatively large distances defined by Van der Waals forces (typically 3-5 Å) are replaced by significantly shorter covalent bonds (e.g., C-C bond, approx. 1.5 Å). This molecular rearrangement — the replacement of long-range VdW distances by short-range covalent bonds — inevitably leads to a significant reduction in the total volume occupied by the molecules. Polymerization shrinkage occurs. The more monomers are incorporated into the polymer network, the more the material contracts. In typical acrylate systems, this can be 5-15%. This shrinkage creates high internal stresses in the material, which can lead to warping, microcracks, and birefringence[14]. The decrease in volume means an increase in density. This increase in density is the primary physical mechanism that links the polymerization kinetics to the optical properties of the material[17].

The refractive index is not an independent material property, but is determined by the electronic structure (polarizability) and density of the material. The Lorentz-Lorenz equation is the fundamental physical law that quantifies this relationship for dielectric materials. It reads:

$$\frac{n^2-1}{n^2+2} = K \cdot \rho \qquad (8)$$

Here, $n$ is the refractive index, $\rho$ is the density of the material, and $K$ is a constant that describes the molar refractivity. The latter is a function of molecular polarizability, which can change slightly during the reaction itself but is often approximated as constant.

According to this equation, the refractive index increases with increasing density, which is also observed in photopolymerization.

## 2. Experimental approach

*2.1 Aim of the experiment*

The primary goal of the experimental investigation is to quantitatively determine the temporal and spatial development of the refractive index during single-photon polymerization. Since the local light dose directly influences the density and thus the refractive index of the polymer, a thorough understanding of this relationship is essential for the precise manufacture of micro-optical components, especially GRIN optics. For this purpose, an experimental setup is used that records both the absolute change in refractive index by means of total reflection at a prism interface and the formation of complex phase gratings in the volume by analyzing diffraction patterns. In addition to characterizing kinetic effects such as oxygen inhibition and reaction rate, the experiment serves to calibrate and validate a simulation model. This model should enable the prediction of optical properties as a function of process parameters, thus creating the basis for an optimized exposure strategy in 3D printing of micro optics.

*2.2 Experimental setup*

Figure 1 shows both a schematic and experimental representation of the experimental setup. The central component is a prism onto whose surface a liquid polymer (PR48) that can be cured by UV polymerization (1-photon process) has been applied. A laser beam is focused on the interface between the sample and the prism (n=1.78). Due to the focusing, light rays from a specific angle range hit the sample. These light rays are either totally reflected or coupled out of the prism according to the condition for total reflection. The critical angle (acceptance angle) at which total reflection occurs is given by:

$$\theta_c = arcsin \frac{n_{polymer}}{n\_prism} \tag{9}$$

By determining the critical angle, the refractive index of the sample can be inferred. If the totally reflected light is now recorded by a camera, the position of the edge of the light-dark transition can be used to determine the total reflection angle and thus directly the local refractive index of the sample (averaged over the area of the focus point of the laser beam – here diameter approx. 2 μm).

For quantitative evaluation, the system (or the light-dark edge) must first be calibrated using samples with a known refractive index. A range of commercial index-matching gels were used for this purpose. In principle, the refractive index of the sample can be determined at the focus point of the laser. To enable spatially resolved measurement, the prism is moved relative to the laser beam using an x-y displacement table. This ultimately provides a two-dimensional measurement of the refractive index distribution on the surface of the sample with a resolution of 2 μm[18].

In order to investigate the change in refractive index during polymerization, a UV exposure mask is projected from below through the prism into the polymer (Wintech4500 / DLP projection system). The exposure mask is made up of alternating on/off pixel rows. This projects a striped UV pattern into the polymer, which leads to striped curing of the polymer. In detail, an "on row" consists of two pixel rows of the DLP projector. Thereby, a sharp image of the pixels is projected onto the polymerization plane, each individual pixel measuring 40μm x 40μm (including the dead zone between pixels). This results in alternating illuminated and unilluminated stripes with a width of 80μm. It should be noted that a single pixel does not allow for homogeneous illumination, as a non-reflective circle is present in the center of a pixel as an additional dead zone due to manufacturing reasons. This also has an influence on the resulting polymerization.

Photo-curing in the polymer results in both a molecular rearrangement, which changes the molar refractive index, and physical shrinkage, which also changes the refractive index. Together, these two effects lead to an increase in the refractive index in the illuminated areas. This increase increases with increasing exposure time. In addition, due to the absorption of UV light in the polymer in the direction of propagation, the UV dose and thus the change in the refractive index decreases. Ultimately, this results in the formation of a complex phase grating in the volume of the polymer. If, in addition to the incoherent UV light, another coherent laser beam (laser 2) is directed through the polymer in parallel, it is diffracted according to the refractive index profile that develops over time in the polymer. This can be recorded and evaluated with the aid of another camera above the prism.

Figure 1b shows the corresponding experimental setup. The bottom image shows the projected pixel pattern, including superposition with the diffraction laser. In order to enable a spatially resolved investigation of the diffraction behavior, the laser beam was split into two spots.

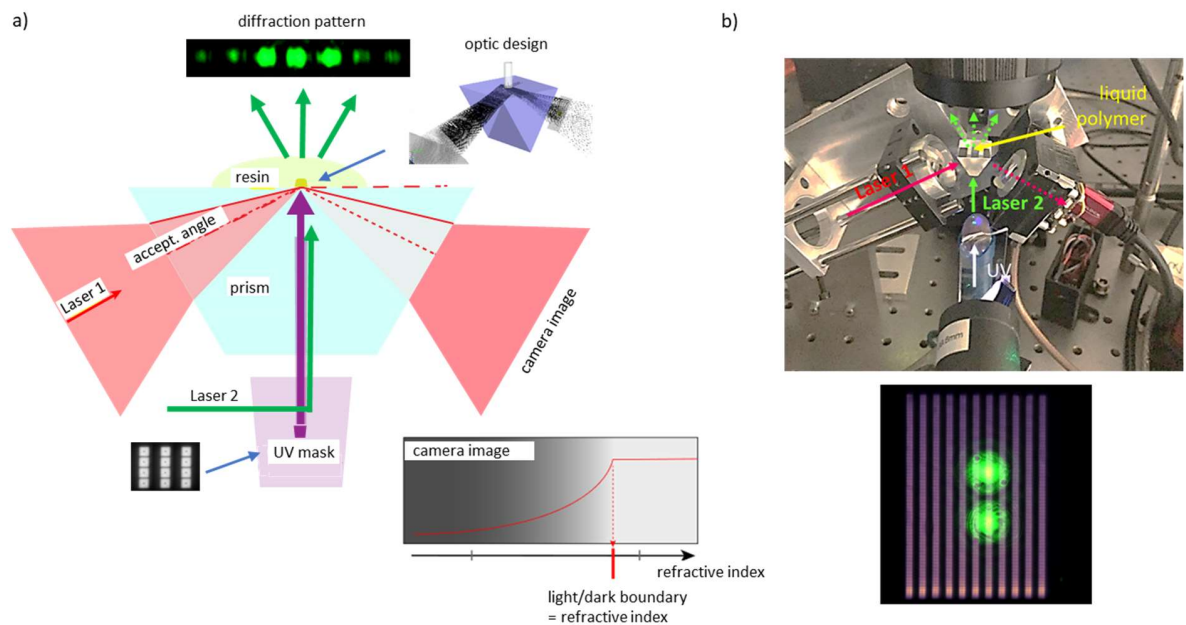

**Fig. 1** Schematic representation of the measuring principle and experimental setup

*2.3 Material used*

Autodesk's Resin PR48 Clear was used as the polymer, as it is an open source 3D printing resin. The main components of this resin are two oligomers and a reactive diluent. Sartomer SR 494 LM (39.776%) and Allnex Ebecryl 8210 (39.776%) are used as oligomers. Rahn Genomer 1122 (19.888%) serves as a reactive diluent and is also incorporated into the polymer chain. This means that the corresponding copolymer is formed from both the oligomers and the reactive diluent. The reactive diluent Genomer can only be cross-linked at the carbon double bond at its end, so that although the chain continues to grow in the copolymer, no branching can take place. In contrast, Sartomer enables cross-linking of several chains via carbon double bonds at all four ends of the molecule. In addition, the photopolymer also contains Esstech TPO+ as a photoinitiator (0.4%) and Mayzo OB+ as a UV blocker (0.16%).

Based on the skeletal formulas for the respective components of the resin, the molar mass, and the density, the theoretical refractive index of PR48 can be estimated to $n = 1.47$, which corresponds well with the experimental data. It should be noted here that the calculation of the molar refractive index leads to a reduction in the refractive index during photopolymerization, since the carbon double bonds lead to a higher molar refractive index compared to the single bonds. However, this reduction in the refractive index is significantly outweighed by the increase in the refractive index due to the increase in density. Ultimately, the refractive index increases by approximately $\Delta n = 0.04$.

The penetration depth defined by the Jacobs model is $D_p = 64\mu m$ for this material, and the critical dose for 20mW/cm² is $E_c = 17.3 \ mJ/cm^2$.

*2.4 Experimental results*

2.4.1 Change in refractive index during UV polymerization at the prism/polymer interface

Figure 2 shows a measurement of the change in the refractive index in the polymer due to UV irradiation (10mW/cm²) onto the surface. The experiment was repeated three times. In addition to the individual measurement curves, the figure also shows the mean value and standard deviation. Irradiation took place from time $t = 5s$. As can be seen, the increase in the refractive index does not occur instantaneously. This is attributed to the oxygen inhibition mentioned above. After approximately 6 seconds, there is a sharp increase in the change in refractive index within a few seconds, which is attributed to the second phase of the polymerization kinetics described above: autoacceleration. This then transits into phase 3 (vitrification) as the reaction time progresses. The course of the refractive index change thus reflects the S-curve of the polymerization kinetics described above.

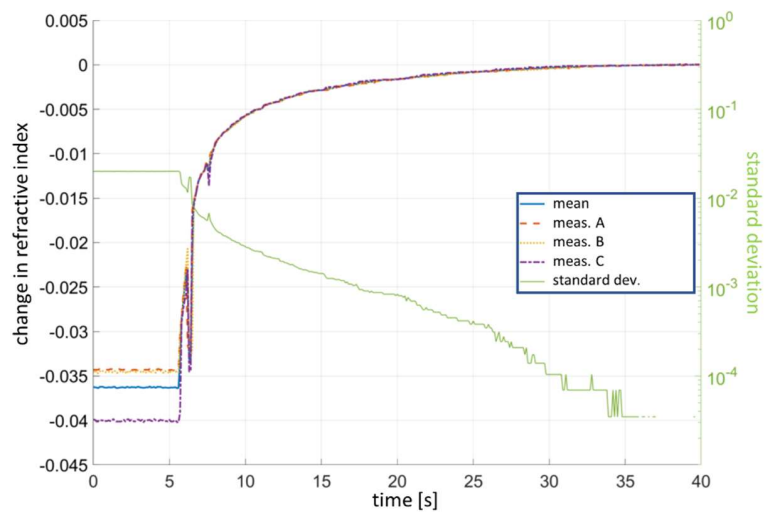

**Fig. 2** Temporal refractive index development during continuous, sustained irradiation at 10 mW/cm².

## 2.4.2 Evaluation of diffraction during photopolymerization

As mentioned above, the laser is split into two light paths so that the diffraction by the phase grating can be measured at two locations simultaneously during polymerization. Figure 3a) shows a typical development of the diffraction patterns. The laser spots were positioned in the center of the UV mask image. At the start time $t = t_0 = 0$, there is a homogeneous refractive index distribution throughout the polymer, so that the two laser beams pass through the sample without diffraction. As time progresses, both laser spots produce approximately the same diffraction patterns, with slight differences in intensity. One reason for this is that the two laser beams were not split 50:50.

As time progresses, it becomes apparent that higher orders occur first (e.g., 0th, 1st, and 2nd order at $t = t_1$), which can be extinguished again depending on the phase development (e.g., 0th order at $t = t_2$). Basically, each diffraction order has a substructure of 3 maxima and 2 minima (see white arrow at $t = t_1$), which, however, is not always visible due to blurring. Once the sample is completely cured, the diffraction pattern freezes accordingly. If the diffraction pattern is then recorded with an increased exposure time (see Fig. 3b), it becomes apparent that the phase grating ultimately diffracts in 2 dimensions.

Overall, it can already be seen in Fig. 3a) and 3b) that this is a complex diffraction structure. This becomes more complicated when looking at the edge areas of the diffraction grating. In Figure 3c), the spots of the diffraction lasers were directed through the upper and middle areas of the grating. The diffraction structures of the respective spots differ significantly. The difference lies not in the order in which the respective spots occur, but in the time at which they appear. This means that the polymerization kinetics is reduced in the edge areas. Two effects are assumed to be possible causes for this. On the one hand, photopolymerization is an exothermic process. This means that during photopolymerization, energy is released as heat, which leads to an increase in temperature. Thermal imaging measurements have shown that, depending on the size of the mask, these temperature increases are in the degree range. The increased temperature leads to an increase in the mobility of the oligomers and thus in the reaction rate. Towards the edge area, there is a corresponding gradient in the temperature profile and thus a decrease in the reaction rate. Another point is that the curing speed also depends on the number of active neighboring pixels. It has been demonstrated that the number of pixels has a direct influence on the curing depth and thus also on the reaction kinetics[19]. Since corresponding neighboring pixels are missing in the edge area, this also leads to a lower reaction speed.

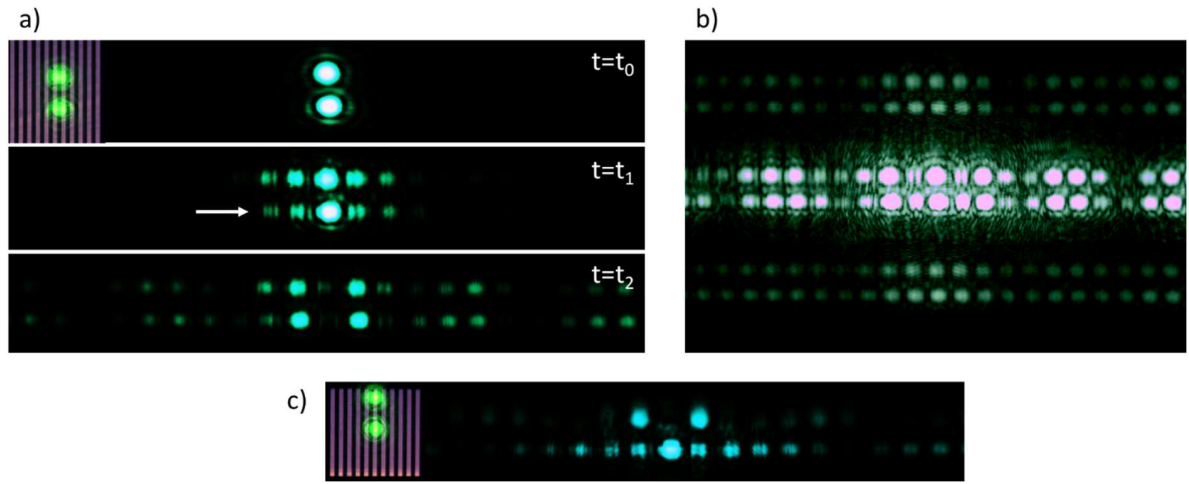

**Fig. 3** Typical change in diffraction patterns over time

To evaluate the diffraction images recorded at 15 Hz, the respective diffraction orders of the two laser spots were separated into two images. Subsequently, conventional image processing algorithms were used to evaluate the intensity of each diffraction order over time. Figure 4 shows a typical curve for the 0th to 3rd (positive) diffraction orders for a laser spot. At time t=0, the UV mask illumination was activated. After approximately 1.2 seconds, a change in the diffraction structure becomes apparent. The intensity of the 0th order decreases. Correspondingly, the intensity of the 1st and 2nd diffraction orders increases due to the redistribution of energy. The 3rd diffraction order follows with a slight delay. As time progresses, the phase shift between the unexposed and exposed areas continues to increase, ultimately causing the intensity of the 0th diffraction order to pass through a minimum and approach a maximum. At the same time, the intensities in the other diffraction orders exhibit the opposite behavior. If an ideal phase grating were present, as described, for example, by the Raman-Nath theory, the behavior for all diffraction orders would repeat oscillatory after reaching the maximum (0th order) or the minimum (higher orders) again. However, after approximately 3 seconds, a deviation from the expected behavior occurs. The second diffraction order no longer shows a symmetrical curve. Similarly, the third diffraction order at 5 seconds, the first diffraction order at 7 seconds, and the zero diffraction order at 9 seconds show an asymmetrical curve. To explain this behavior, it is assumed that the following four effects occur during photopolymerization:

1. The formation of polymer chains not only changes the refractive index, but also the scattering properties of the polymer. It is assumed that scattering increases with increasing polymerization due to the increasing size of the polymers. This leads to a decrease in the total absorbed diffraction intensity over time.
2. Although mask exposure creates a line pattern in the polymer, light scattering also occurs in the unexposed areas between the lines. This also leads to (slight) curing in these areas. This scattering effect increases with increasing cured layer thickness.
3. The radicalized oligomers are not limited to the exposed area in terms of their mobility. They also diffuse into the unexposed area, where they also enable polymerization.
4. Polymerization is accompanied by shrinkage in the polymer. This can be up to 10%.

Points 2-4 are considered essential for the temporal course of the diffraction intensities. Points 2 and 3 imply that the difference in refractive index between the exposed (polymerized) and unexposed (non-polymerized) areas is not constant, but decreases over time. This leads to a dynamic change in the temporal progression of the diffraction orders. Point 4 also results in a dynamic change in the refractive index progression, as increasing shrinkage causes the lattice constant of the diffraction grating to change over time, resulting in a shift of the diffraction orders.

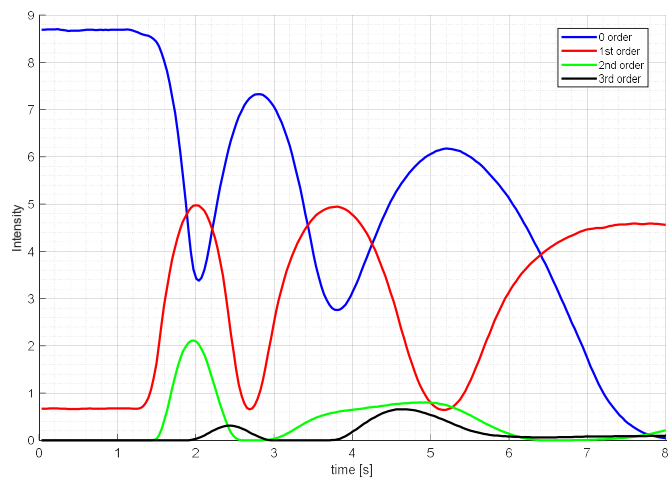

**Fig. 4** Intensity curve of the 0th to 3rd diffraction orders versus time

In addition, the diffraction patterns can be used to draw conclusions about the kinetics during polymerization. For example, the oxygen inhibition time can be determined. Figure 5a shows the difference between the intensity of the 0th diffraction order at time $t = 0$ and the intensity of the 0th order at time $t > 0$ for different irradiance levels ($\Delta I = I_{0-ord\ ,t>0} - I_{0-ord\ ,t=0}$). In principle, a decrease in the difference in intensity can be seen for all irradiance levels. This corresponds to the redistribution of light intensities to other orders. It can also be seen that for lower irradiance levels, a change in the diffraction pattern of the 0th order occurs later than for higher irradiance levels. Assuming that the oxygen concentration in the polymers is constant for all experiments, this shows that higher irradiance levels break down oxygen inhibition more quickly. To quantify this, Figure 5b) shows the inflection point of the respective curve (= oxygen inhibition is broken down) plotted semi-logarithmically against the irradiance. The higher the light intensity, the exponentially shorter the waiting time until the start of polymerization. Furthermore, it can be seen that the slope of the curves increases with increasing irradiance. As can be seen in Figure 5c), this results in a linear relationship.

The different dependencies of the inhibition time and the propagation rate on the irradiance observed in the experimental data can be attributed to fundamental kinetic and photochemical principles. The exponential decrease in oxygen inhibition time results primarily from competitive kinetics between photochemical oxygen consumption and the continuous post-diffusion of atmospheric oxygen into the reaction volume. At low intensities, the diffusion process is dominant and disproportionately prolongs the time required to reach the critical energy $E_c$ necessary for polymerization, as new inhibitor is continuously supplied to the system. However, as the irradiance increases, the system transits to a reaction-controlled state in which the radical formation rate far exceeds the post-diffusion and the inhibition barrier is efficiently broken. This transition, coupled with the exponential attenuation of light in the material according to Beer-Lambert's law, manifests itself in the exponential characteristic of the inhibition time.

In contrast, after the inhibition phase has been overcome, the propagation rate follows a linear relationship that is directly derived from the equivalence law of photochemistry. Since the initiation rate is proportional to the absorbed photon flux density, an increase in irradiance leads to a direct, linear increase in the concentration of active radicals. Since the chain growth rate depends stoichiometrically on the number of these active chain ends and there are no autocatalytic feedback effects, this results in the observed linear scaling of the reaction rate with the light power supplied.

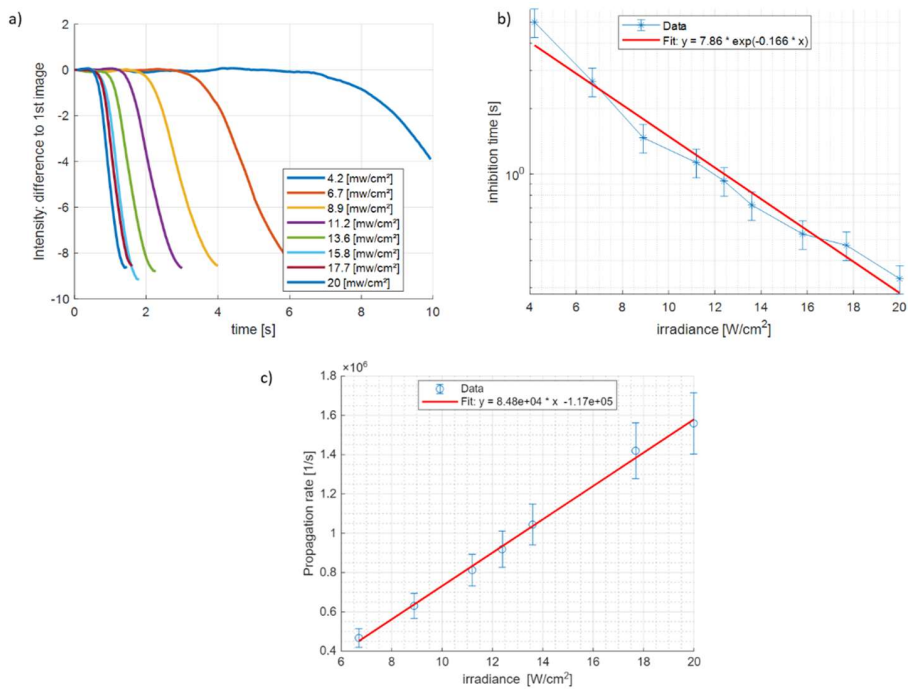

**Fig. 5 a)** Intensity curve of the 0th to 3rd diffraction order versus time; b) Dependence of the inhibition time on the irradiation intensity; c) Influence of the irradiation intensity on the propagation rate.

## 3. Modeling and simulation

The diffraction intensities are a direct consequence of a periodic change in the refractive index and thus indirect evidence of the polymerization taking place. For the realization of optical microelements using single-photon polymerization, an understanding of the resulting refractive index distribution is therefore essential. Ultimately, the goal is to find a descriptive model for photopolymerization that establishes the relationship between the optical effect of the optical element and the process parameters. For this reason, a model that (partially) replicates the diffraction experiments will be presented below. One challenge here is the complexity of photopolymerization due to material shrinkage, light scattering, and radical diffusion. This causes a change in the refractive index not only in the exposed part, but also in the unexposed part. In order to reduce this complexity as a first step, it is assumed in the following that photopolymerization takes place only in the exposed areas, neglecting shrinkage.

*3.1. Basics for simulation and model structure*

The diffraction pattern is simulated in accordance with Fourier optics in the Fraunhofer approximation. According to this, the following applies to the field in the observation plane:

$$U_2(x,y) = = \frac{e^{jkz}}{j\lambda z} exp\left[j\frac{k}{2z}(x^2+y^2)\right] \times \mathcal{F}\{U_1(x,y)\} \tag{10}$$

Where $\lambda$ is the wavelength, $z$ is the distance from the camera to the diffraction location, and $U_1$ is the field at the diffraction location. Accordingly, the following applies to the intensity distribution in the observation plane:

$$I_2(x,y) = \left(\frac{1}{\lambda z}\right)^2 (\mathcal{F}\{U_1(x,y)\})^2 \tag{11}$$

and for diffraction plane one gets:

$$U_1(x,y) = U_0 * e^{i\phi(x,y)} \tag{12}$$

where the phase is given by:

$$\phi = \frac{2\pi}{\lambda} \Delta n(x,y,z,t)\, z(x,y,t) \tag{13}$$

Thus, the intensity depends on the difference in refractive index between the exposed and unexposed areas and the height of the cured area. Formula (13) also directly shows the challenge for the simulation. The intensity at the observation point results from the superposition of all electromagnetic waves coming from the volume of the polymer and interfere at the observation point. Due to location-dependent polymerization, these waves undergo a location-dependent phase shift, as the refractive index changes locally over time within the volume. Accordingly, $\Delta n$ is a function of the three spatial coordinates and time. The curing front migrating $z(x,y,t)$ into the polymer is a function of (x,y) and time.

Approximation for the refractive index function $\Delta n(x,y,z,t)$:
In principle, $\Delta n$ can be defined as the difference between the refractive index of the polymerized state and that of the liquid polymer. During the transition from the liquid state to the solid state, the refractive index exhibits an S-shaped curve, as shown in Figure 2. The measurement data from Figure 2 was recorded at the interface between the prism and the polymer using flat projector illumination at the measurement location. This curve therefore represents the fastest possible change in refractive index during polymerization. The development of the refractive index profile in the z-direction is decisive for the formation of the intensities of the diffraction orders. This is not accessible in this way, but it can be assumed that it also has an S-shaped curve equivalent to Figure 2. However, it must be taken into account that, in accordance with Lambert Beer's law, the intensity of the UV projection decreases with increasing depth, which leads to reduced curing. The thermal properties in the polymer also differ from those at the prism interface.

Further information is available in the form of the time that elapses until no further change in the intensities of the diffraction orders can be detected. The entire structure is polymerized, i.e., the end point of the S-curve has been reached. In order to obtain a model for $\Delta n(x, y, z, t)$, the measured S-curve is stretched to this point in time. This new S-curve is then used for modeling $\Delta n(z, t)$. The spatial coordinate $z$ is taken into account by assuming that a curing front migrates into the polymer with the position $z_{cured}(t)$. Only $z < z_{cured}$ is used for $\Delta n(z, t)$. Outside of this, $\Delta n(z, t) = 0$ applies.

In the simulation, time is traversed in discrete steps. The refractive index difference applicable for the respective time step is then given accordingly by $\Delta n(z, t)$ and is used as first input to calculate the time dependent phase.

Approximation for the shape function $z(x, y, t)$:

Photopolymerization occurs due to mask-shaped exposure. This must be mapped as a shape factor in order to derive the height profile $z(x, y, t)$ (=curing front) for the phase calculation. To do this, the shape of cured pixels was first recorded using a white light interferometer (see Fig. 6a). It can be seen that the dead zone inherent in DLP projectors between the individual pixels causes reduced curing of the polymer in these areas. Furthermore, the figure shows that the circular dead zone in the center of a pixel also leads to reduced curing. However, this is significantly smaller than the dead zone around the pixel. The reason for this is assumed to be that surrounding areas also scatter UV light into the dead zone in the center of the pixel during curing.

The profile of a single cured pixel was then translated into a normalized shape factor (Figure 6b). For the central circular "dead zone", this results in a reduction in height of approximately 40% compared to the plateau. The model neglected the fact that the pixels at their corner points cause less curing than at the midpoints of the side edges. The shape factor for a single pixel is thus surrounded by a complete dead zone (height $z = 0$ / blue in Fig. 6b). The shape generated in this way was finally normalized to 1 and duplicated to produce a corresponding grid (see Fig. 6c).

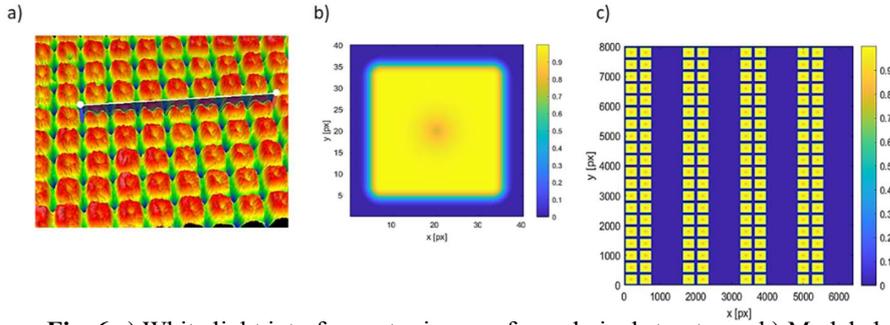

**Fig. 6** a) White light interferometer image of cured pixel structures. b) Modeled shape of a single pixel and duplication of the single pixel to form the projection mask.

The 3D function obtained in this way, normalized in terms of height, is then converted into a 3D shape factor that increases over the simulation time and describes the movement of the curing front. However, due to kinetics, the curing front does not move linearly into the polymer during polymerization. In order to be able to assign a specific height to a specific point in time, an S-shaped curve of height over time was also assumed. The maximum height after complete polymerization can be taken from the measured layer thickness of the cured samples and used for scaling. Figure 7 shows the modelled heights of the cured profile at different points in time.

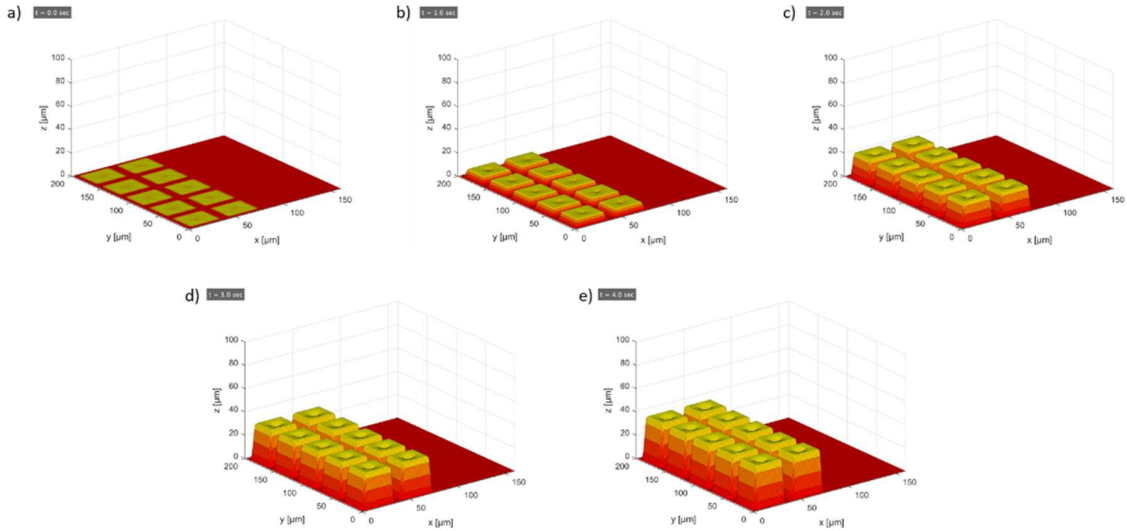

**Fig. 7** Section of the 3D form factor or curing front at different points in time: a) $t = 0s$ ; b) $t = 1s$ ; c) $t = 2s$; d) $t = 3s$ and e) $t = 4s$

In the simulation, based on the two models for $\Delta n(x, y, z, t)$ and $z(x, y, t)$, the phase and thus, as described above, the intensity of the individual diffraction orders are simulated as a function of time.

## 3.2. Simulation results and discussion

Figure 8 shows the simulated intensities of the various diffraction orders at any given time. The upper figure in 8a shows a linear representation and the lower figure a logarithmic representation of the intensity. The logarithmic representation shows that the distance between the diffraction structures is comparable to the distance between the experimentally determined diffraction intensities (compare Fig. 3a). Furthermore, it can be seen that the simulation also reflects the experimentally observed triple substructure. In order to investigate the cause of the triple substructure, the pixel shape, pixel size, etc. were varied. This showed that the reason for the substructure lies in the dead zone of the exposure around the pixel (shown in blue in Figure 6b). If this dead zone is removed, the substructure transits into a single intensity peak.

Figure 8b shows a 2D representation of the simulated diffraction intensities. This also corresponds to the experimental observations (see Fig. 3b). The distance in the x-direction between the diffraction orders is smaller than the distance in the y-direction. This is because the diffraction in the x-direction is caused by the distance between the on/off pixel columns (period: 80 µm). The reason why a diffraction pattern is created in the y-direction is due to the subdivision of the diffracting structure into pixels in the y-direction (period 40 µm). The smaller distance causes stronger diffraction, which leads to a larger distance between the diffraction signals in the y-direction.

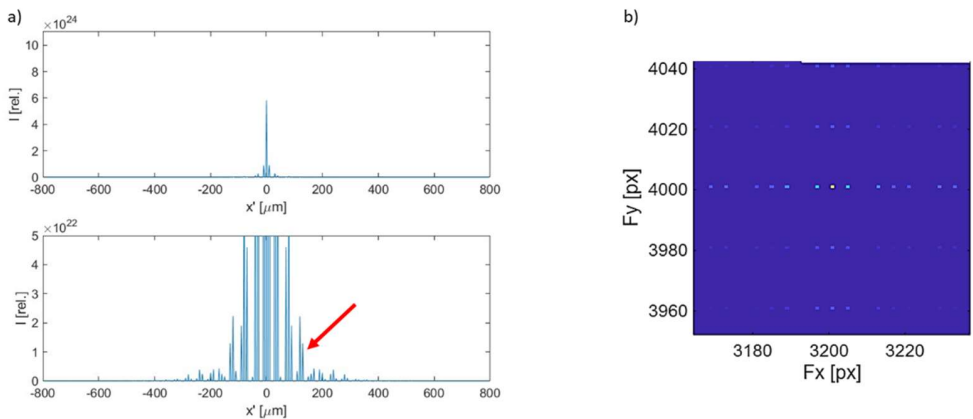

**Fig. 8** a) Profile section of the intensity distribution of the diffraction orders at any given time; b) 2D representation of the intensity distribution.

Figure 9a shows both the measured (solid line) and simulated (dashed line) intensity of the 0th, +1st, and +2nd diffraction orders. The assumed behaviour $\Delta n$ of is also shown. It can be seen that the simulation accurately reproduces the intensity curve of the 0th and 1st diffraction orders up to approx. 6 seconds. This corresponds to approximately 1/2 of the curing time. The 2nd diffraction order, on the other hand, is only well reproduced up to approximately 3 seconds. In the experiment, a deviation from the conventional expected behaviour can be seen for the 0th and 1st diffraction orders from approximately 6 seconds, and for the 2nd diffraction order from approximately 3 seconds. The assumptions made regarding an S-shaped increase of $\Delta n$ does not longer apply for this regime. In the experiment, the diffraction orders disappear more slowly than predicted by the simulation. As mentioned above, one possible reason for this might be that curing also takes place in the unexposed areas. In order to investigate this possibility in principle, a further simulation (Figure 9b) assumed that the refractive index also increases in the unexposed area. Ultimately, these unexposed areas also polymerize completely, which is consistent with the experimental observations. For the model, it is therefore assumed that the refractive index in the unexposed area increases more slowly and with a delay. This results in the curve shown in Figure 9b) for $\Delta n$. If we consider the simulation of the second diffraction order in this case, for example, the first maximum at 2 seconds and the asymmetrical curve from 3-5 seconds are well reproduced. However, there is another peak at approximately 6 seconds, which cannot be observed in the experiment. Nevertheless, this investigation shows that a more complex, time-varying refractive index distribution over the entire range must be assumed for a simulation of the entire time regime.

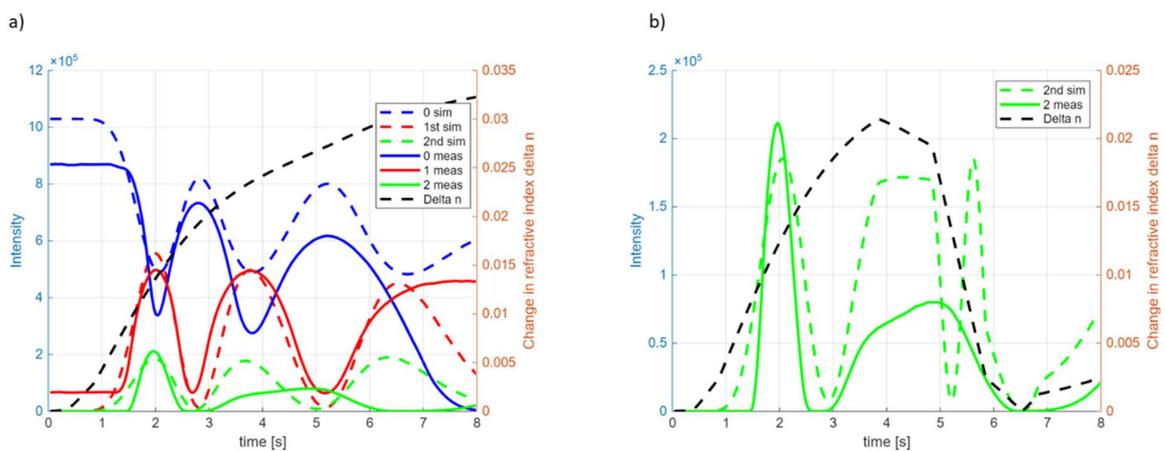

**Fig. 9** a) Comparison of the measured intensities with the simulated intensities for the 0th, 1st, and 2nd diffraction orders, as well as the behaviour of the assumed refractive index curve; b) improved model for $\Delta n$, based on an assumed change in the refractive index for the unexposed area as well.

4. Summary and outlook

In this work, the dynamics of the refractive index change during the 1-photon polymerization of acrylate resins was experimentally investigated and modelled. Using a coupled setup of total reflection and diffraction, it was shown that the refractive index follows a significant S-shaped increase that strongly depends on the local irradiance. The oxygen inhibition time showed an exponential decrease with increasing intensity.

Analysis of the diffraction patterns provided deeper insights into the micro structuring. It was found that the "dead zones" between the DLP pixels lead to characteristic substructures in the diffraction orders. A key finding is the evidence that polymerization is not sharply limited to the exposed areas. Both experimental data and the Fourier optics-based simulation model show that scattering and radical diffusion cause delayed curing even in nominally dark areas. Over time, this leads to a decrease in the refractive index contrast and asymmetrical behaviour of the diffraction intensities.

Future work will investigate the influence of exothermic heating on the refractive index gradient in more detail, as temperature changes affect the mobility of the oligomers and thus the kinetics in peripheral areas. In addition, an extension of the simulation model is necessary, which dynamically couples the volume change (shrinkage) and the depth dependence of curing (Beer-Lambert law). Such an extended model could be used to develop exposure strategies (e.g., gray value modulation) that compensate for unwanted curing in dark areas, thus enabling defined optical functionality of printed microelements.

*Disclosures*

The authors declare that there are no financial interests, commercial affiliations, or other potential conflicts of interest that could have influenced the objectivity of this research or the writing of this paper.

*Code, Data, and Materials Availability*

The archived version of the code described in this manuscript and a recorded diffraction pattern can be freely accessed through GitHub (https://github.com/AndreasHeinrichAalen/CodeFFT)

*Acknowledgments*

The authors would like to express their special thanks to Anne Harth for the detailed discussions and her valuable contributions. Furthermore, the authors would like to thank the German Research Foundation for its financial support (project InProGRIN - project number HE 3533/13-1). DeepL was used for language and grammar clean-up.

**Andreas Heinrich** is a full professor at the Aalen University. He received his Diploma in physics from the Technical University of Munich in 1998, and his PhD degree in experimental physics from the Technical University of Munich in 2001. He completed his postdoctoral lecture qualification at the University of Augsburg in 2007. He is the author of more than 100 journal papers and has written four book chapters. His current research interests include 3D printing of optics, behaviour of polymers in electric fields, refractive index dynamics in polymers, Nano-Imprint Lithography and Physical Informed Neural Networks.

Biographies and photographs for the other authors are not available.

**Caption List**

Fig. 1 Schematic representation of the measuring principle and experimental setup

Fig. 2 Temporal refractive index development during continuous, sustained irradiation at 10 mW/cm².

Fig. 3 Typical change in diffraction patterns over time

Fig. 4 Intensity curve of the 0th to 3rd diffraction orders versus time

Fig. 5 a) Intensity curve of the 0th to 3rd diffraction order versus time; b) Dependence of the inhibition time on the irradiation intensity; c) Influence of the irradiation intensity on the propagation rate.

Fig. 6 a) White light interferometer image of cured pixel structures. b) Modeled shape of a single pixel and duplication of the single pixel to form the projection mask.

Fig. 7 Section of the 3D form factor or curing front at different points in time: a) ) $t = 0s$ ; b) $t = 1s$ ; c) $t = 2s$; d) $t = 3s$  and e) $t = 4s$

Fig. 8 a) Profile section of the intensity distribution of the diffraction orders at any given time; b) 2D representation of the intensity distribution.

Fig. 9 a) Comparison of the measured intensities with the simulated intensities for the 0th, 1st, and 2nd diffraction orders, as well as the behaviour of the assumed refractive index curve; b) improved model for $\Delta n$, based on an assumed change in the refreactive index for the unexposed area as well.